\documentclass[10pt, a4paper]{article}
\usepackage[final]{lrec2026} % this is the new style
\usepackage{
url,
comment,
placeins,
booktabs,
multirow,
graphicx,
array,
pgfplots,
xcolor,
pgfplotstable,
caption,
subcaption,
amsmath,
amssymb,
amsfonts,
algorithmic,
textcomp,
lineno,
CJKutf8,
enumitem,
caption,
}

\title{TG-ASR: Translation-Guided Learning with Parallel Gated Cross Attention for Low-Resource Automatic Speech Recognition}

\name{
\begin{tabular}{c}
Cheng-Yeh Yang\textsuperscript{1}, Chien-Chun Wang\textsuperscript{1},
Li-Wei Chen\textsuperscript{3}, Hung-Shin Lee\textsuperscript{3}, \\
Hsin-Min Wang\textsuperscript{2},
and Berlin Chen\textsuperscript{1}
\end{tabular} \\
}

\address{
\textsuperscript{1}Dept. Computer Science and Information Engineering, National Taiwan Normal University, Taiwan \\
\textsuperscript{2}Institute of Computer Science, Academia Sinica, Taiwan \\
\textsuperscript{3}United Link Co., Ltd., Taiwan
}

\abstract{
Low-resource automatic speech recognition (ASR) continues to pose significant challenges, primarily due to the limited availability of transcribed data for numerous languages.
While a wealth of spoken content is accessible in television dramas and online videos, Taiwanese Hokkien exemplifies this issue, with transcriptions often being scarce and the majority of available subtitles provided only in Mandarin.
To address this deficiency, we introduce TG-ASR for Taiwanese Hokkien drama speech recognition, a translation-guided ASR framework that utilizes multilingual translation embeddings to enhance recognition performance in low-resource environments.
The framework is centered around the parallel gated cross-attention (PGCA) mechanism, which adaptively integrates embeddings from various auxiliary languages into the ASR decoder.
This mechanism facilitates robust cross-linguistic semantic guidance while ensuring stable optimization and minimizing interference between languages.
To support ongoing research initiatives, we present YT-THDC, a 30-hour corpus of Taiwanese Hokkien drama speech with aligned Mandarin subtitles and manually verified Taiwanese Hokkien transcriptions.
Comprehensive experiments and analyses identify the auxiliary languages that most effectively enhance ASR performance, achieving a 14.77\% relative reduction in character error rate and demonstrating the efficacy of translation-guided learning for underrepresented languages in practical applications.
\\ \newline \Keywords{Low-resource automatic speech recognition, Taiwanese Hokkien, translation-guided learning, parallel gated cross attention, multilingual auxiliary language integration}
}

\pgfplotsset{compat=1.18}

\begin{document}

\maketitleabstract

\section{Introduction}

While multilingual corpora and large-scale speech datasets have significantly advanced automatic speech recognition (ASR) \cite{ardila2020,pratap2020,wang2021}, many regional and endangered languages remain critically underrepresented \cite{besacier2014,yeroyan2024,bartelds2023}.
The scarcity of transcribed speech, linguistic resources, and standardized evaluation benchmarks continues to hamper the development of ASR for low-resource languages.
ASR systems trained on high-resource languages, such as Mandarin or English, often exhibit difficulties in generalizing to typologically distinct and data-sparse languages, leading to unstable convergence and diminished transcription accuracy.

\begin{figure}[t]
\centering
\includegraphics[width=1.0\linewidth]{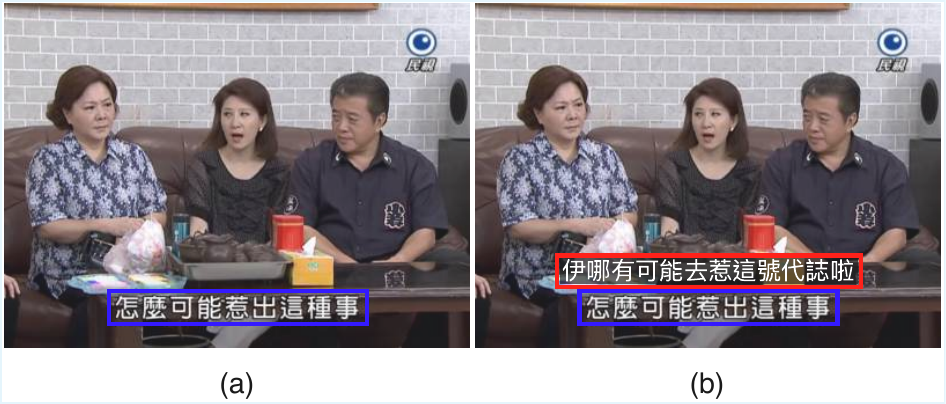}
% \vspace{-15pt}
\caption{
Illustration of the Taiwanese Hokkien drama subtitles.
(a) A scene with spoken Taiwanese Hokkien and existing Mandarin subtitles enclosed in a blue box\protect\footnotemark.
(b) The outcome of our framework, showing automatically generated Taiwanese Hokkien subtitles, highlighted with a red box, aligned with the Mandarin subtitles.
}
% \vspace{-10pt}
\label{fig:scenario}
\end{figure}

In Taiwan, Taiwanese Hokkien exemplifies a significant challenge in the realm of speech processing.
Despite the availability of numerous Taiwanese Hokkien TV dramas and online videos, most subtitles are predominantly written in Mandarin, which results in the speech of Taiwanese Hokkien being largely untranscribed.
This discrepancy not only restricts the accessibility of Taiwanese Hokkien-language media but also threatens the preservation of cultural heritage.
To address this issue, we propose the development of an ASR system capable of generating Taiwanese Hokkien transcriptions that are aligned with the existing Mandarin subtitles.
The primary objectives of our system are to enhance bilingual accessibility, facilitate subtitle generation, and contribute to the preservation of the language.
Figure~\ref{fig:scenario} illustrates this context, where (a) depicts a typical scene from a Taiwanese Hokkien drama featuring only Mandarin subtitles, while (b) showcases our envisioned outcome, which includes subtitles in Taiwanese Hokkien.
To support this initiative, we have constructed and released a novel corpus, \textbf{YT-THDC} (\textbf{Y}ou\textbf{T}ube-sourced \textbf{T}aiwanese \textbf{H}okkien \textbf{D}rama \textbf{C}orpus), comprising approximately 30 hours of Taiwanese Hokkien drama speech that is meticulously aligned with Mandarin subtitles and manually verified Taiwanese Hokkien text transcriptions.
This corpus serves as a valuable benchmark for low-resource ASR research and facilitates a systematic investigation of translation-guided approaches within real-world media contexts.

Traditional approaches to low-resource ASR have explored data augmentation \cite{park2019,ko2015}, multilingual joint training \cite{conneau2021,kannan2019}, and transfer learning from resource-rich languages \cite{watanabe2017,bapna2022}.
However, these methods exhibit limited efficacy when speech data is exceedingly scarce and linguistically divergent from pre-trained languages.
Specific to Taiwanese Hokkien, researchers have also attempted to automatically generate Taigi transcriptions from dramas by aligning the speech with existing Mandarin subtitles using an initial ASR system \cite{chen2020}.
Recent research indicates that integrating auxiliary information, such as translated text, can provide additional semantic supervision, thereby enhancing ASR performance \cite{soky2022,taniguchi2023,rouditchenko2024}.
Motivated by this observation, we investigate how translation embeddings from different \emph{auxiliary languages} can assist ASR in a specific \emph{target language}, Taiwanese Hokkien.

\footnotetext{
The meaning of the subtitle is ``How could he possibly get involved in such a thing?'' in English.
Images were adapted from publicly available content by Formosa Television Co., Ltd. for educational use.}

In this study, we present \textbf{TG-ASR}, an innovative \textbf{T}ranslation-\textbf{G}uided \textbf{A}utomatic \textbf{S}peech \textbf{R}ecognition framework designed for Taiwanese Hokkien drama speech recognition in low-resource scenarios.
Specifically, our framework introduces the Parallel Gated Cross-Attention (PGCA) mechanism, which integrates multilingual translation embeddings extracted from pre-trained language models into the ASR decoder.
PGCA adaptively regulates the contribution of each auxiliary language, facilitating effective cross-lingual semantic guidance while maintaining stable optimization and minimizing language interference.
Through extensive experiments, we analyze the auxiliary languages that most significantly enhance Taiwanese Hokkien ASR performance and demonstrate the practical utility of translation-guided learning in real-world applications.
The main contributions of this study are summarized as follows:
\begin{enumerate}[noitemsep,leftmargin=*]
\item{\textbf{An Innovative Translation-Guided ASR Framework:} We propose TG-ASR, a novel framework designed for low-resource Taiwanese Hokkien drama speech recognition. This framework introduces the PGCA mechanism to flexibly and adaptively integrate multilingual translation embeddings, enhancing recognition performance and training stability.}
\item{\textbf{A New Corpus for Low-Resource Research:} The YT-THDC corpus, a 30-hour collection of Taiwanese Hokkien dramas with corresponding transcriptions in both Taiwanese Hokkien and Mandarin, is released to facilitate future research in low-resource ASR and bilingual subtitle generation.}
\end{enumerate}

\begin{table}[t]
\small
\centering
\setlength{\tabcolsep}{12.5pt}
\begin{tabular}{lcc}
\toprule
\textbf{Split} & \textbf{Duration (hours)} & \textbf{\# Utterances} \\
\midrule
Train & 27.51 & 50,984 \\
Test  & 2.79  & 4,859  \\
\midrule
Total & 30.30 & 55,843 \\
\bottomrule
\end{tabular}
% \vspace{-5pt}
\caption{Statistics of YT-THDC, including total duration and number of utterances in each split.}
\label{tab:corpus}
% \vspace{-15pt}
\end{table}

\section{Corpus: YT-THDC}

The YouTube-sourced Taiwanese Hokkien Drama Corpus (YT-THDC) is a newly constructed dataset specifically designed for the training and evaluation of ASR models in Taiwanese Hokkien.
The corpus comprises numerous drama series sourced from publicly available YouTube videos, encompassing a diverse range of speakers, scenes, and recording environments.
This diversity offers rich linguistic and acoustic variability, closely reflecting authentic usage conditions.
Each video features spoken Taiwanese Hokkien paired with Mandarin subtitles, which serve as loosely aligned references rather than exact transcriptions of the spoken content.
To obtain precise speech–text pairs, the Taiwanese Hokkien speech was initially transcribed using a pre-trained ASR model, with the outputs subsequently manually verified and refined by linguistic experts, ensuring high transcription fidelity suitable for supervised ASR training.
The final corpus consists of approximately 30 hours of speech and over 50,000 utterances, as summarized in Table~\ref{tab:corpus}.
All recordings naturally include multiple speakers, background music, and ambient noise, characteristics that render the corpus particularly valuable for the development of noise-robust and context-aware ASR systems.
YT-THDC is available for research use only and serves as a foundational resource for studying low-resource Taiwanese Hokkien ASR as well as multilingual transfer learning with cross-lingual text supervision.

\begin{figure*}[ht]
\centering
\includegraphics[width=1.0\linewidth]{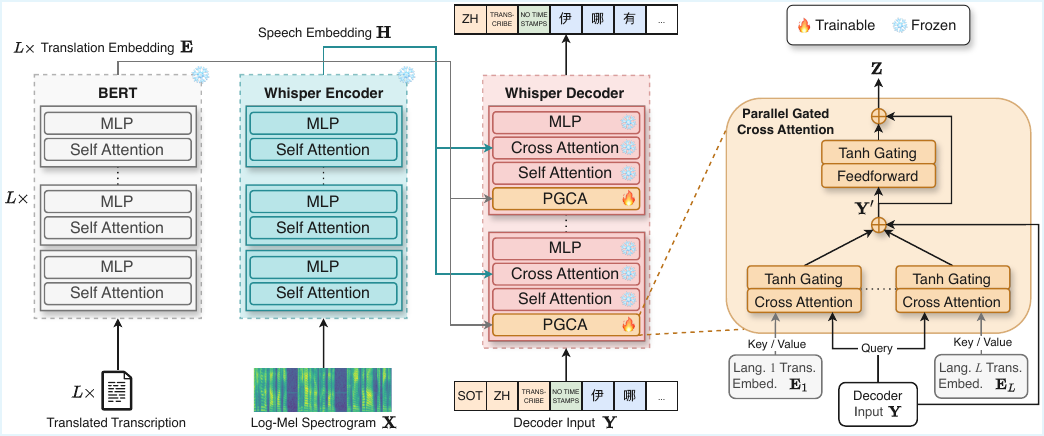}
% \vspace{-15pt}
\caption{The architecture of the proposed TG-ASR, which leverages our novel parallel gated cross-attention (PGCA) mechanism to integrate multilingual translated transcription inputs for improved knowledge transfer in ASR.}
% \vspace{-10pt}
\label{fig:main}
\end{figure*}

\section{Methodology: TG-ASR}

\subsection{Framework Overview}

Figure~\ref{fig:main} illustrates the architecture of our translation-guided framework, TG-ASR, trained in a two-stage process.
In the first stage, the entire Whisper model, where both the encoder and decoder are updated simultaneously.
The Whisper encoder converts input speech $\mathbf{X}$ into acoustic embeddings $\mathbf{H}$, which are subsequently processed by the decoder for transcription and language modeling.
In the second stage, the decoder parameters obtained from the first stage by incorporating our proposed parallel gated cross-attention layers and perform fine-tuning.
During this stage, the Whisper encoder and the decoder parameters from the first stage remain frozen, allowing only the PGCA layers to be updated.
Auxiliary language transcriptions generated by SeamlessM4T \cite{barrault2023} are encoded using a pre-trained Multilingual BERT \cite{devlin2019} to produce multilingual translation embeddings $\mathbf{E}$.
These embeddings are integrated into the decoder via the PGCA layers, where a learnable gating mechanism adaptively regulates the contribution from each auxiliary language.
The resulting weighted embeddings are aggregated and added to the decoder input $\mathbf{Y}$ before being processed through a feedforward neural network (FNN).
This design enables the model to incorporate multilingual semantic guidance while preserving the core decoding process of Whisper and limiting parameter updates exclusively to the PGCA layers.

\subsection{Multilingual Embedding Extraction}

To integrate multilingual semantic knowledge, we leverage multilingual BERT (mBERT) \cite{devlin2019}, a Transformer-based model that generates contextualized word embeddings.
BERT is pre-trained using masked language modeling and next sentence prediction, enabling it to effectively capture linguistic dependencies across multiple languages.
In our framework, we employ mBERT \textsubscript{BASE} with 12 layers, 768 hidden units, and 12 attention heads.
It extracts multilingual embeddings $\mathbf{E}_l \in \mathbb{R}^{T_l \times d}$ from machine-translated auxiliary language transcriptions generated by SeamlessM4T \cite{barrault2023}, where $T_l$ and $d$ denote the number of tokens in auxiliary language $l$ and the embedding dimension, respectively.
These embeddings, which encode cross-lingual semantic cues, are integrated into the Whisper decoder via the PGCA mechanism, thereby enhancing recognition robustness in multilingual ASR scenarios.
Notably, mBERT remains frozen during training, ensuring that multilingual information is consistently leveraged without additional fine-tuning.

\subsection{Acoustic Embedding Extraction}

The Whisper encoder \cite{radford2023} transforms raw speech into high-level representations that are conducive for decoding.
It processes log-mel spectrograms $\mathbf{X} \in \mathbb{R}^{T_s \times F}$ through a convolutional front-end, followed by Transformer encoder blocks, wherein self-attention mechanisms capture both local and global temporal dependencies \cite{vaswani2017}.
Here, $T_s$ denotes the number of time frames, while $F$ indicates the number of mel-frequency bins.
The resultant sequence of contextualized speech embeddings $\mathbf{H} \in \mathbb{R}^{T_s \times d}$ yields a robust representation of the input signal, facilitating accurate and cross-lingual ASR transcription within our framework.
Similar to mBERT, the Whisper encoder remains frozen during the PGCA training stage , leveraging the adapted representations acquired from the initial fine-tuning phase.

\subsection{Multilingual Fusion}

To enhance target language speech recognition, we extend the Whisper decoder \cite{radford2023} with gated cross-attention \cite{alayrac2022} that incorporates translated transcriptions, as illustrated in Figure~\ref{fig:main}.
Our framework integrates auxiliary translation information while preserving the core architecture and functionality of the Whisper decoder blocks.
The original Whisper decoder block consists of self-attention, cross-attention (attending to audio features), and a multi-layer perceptron (MLP).
Inspired by Flamingo \cite{alayrac2022}, we introduce a parallel gated cross-attention mechanism to fuse multilingual embeddings derived from translated transcriptions.
Unlike Whisper-Flamingo \cite{rouditchenko2024}, which employs a single cross-attention module for visual input, our framework introduces multiple parallel cross-attention modules.
Each module independently attends to multilingual embeddings $\mathbf{E}_l$ derived from $L$ translated versions of the auxiliary language transcription, where $L$ denotes the total number of auxiliary languages considered.
Formally, given decoder input $\mathbf{Y} \in \mathbb{R}^{T_y \times d}$ and multilingual embeddings $\{\mathbf{E}_1, \dots, \mathbf{E}_L\}$, where $T_y$ denotes the number of tokens in the decoder input, the PGCA mechanism operates as follows:
\begin{equation}
\mathbf{Y}' = \mathbf{Y} + \sum_{l=1}^{L} \operatorname{tanh}(\alpha_{\operatorname{attn}^{(l)}}) \times \operatorname{attn}(\mathbf{Y}, \mathbf{E}_l, \mathbf{E}_l),
\end{equation}
\begin{equation}
\mathbf{Z} = \mathbf{Y}' + \operatorname{tanh}(\alpha_{\operatorname{FNN}}) \times \operatorname{FNN}(\mathbf{Y}'),
\end{equation}
where $\alpha_{\operatorname{attn}}^{(l)}$ and $\alpha_{\operatorname{FNN}}$ are learnable gating parameters that dynamically regulate the influence of each auxiliary language in conjunction with the feedforward transformation.
In this context, $\operatorname{tanh}$ denotes a gating function that constrains the contribution range, $\operatorname{attn}$ represents the cross-attention mechanism between decoder states and translation embeddings, and $\operatorname{FNN}$ refers to the feedforward neural network that refines the integrated representation.
All gating parameters ($\alpha_{\operatorname{attn}}^{(1)}$, $\dots$, $\alpha_{\operatorname{attn}}^{(L)}$, and $\alpha_{\operatorname{FNN}}$) are initialized to zero to ensure training stability.
The PGCA modules are strategically positioned at the outset of each Whisper decoder block to seamlessly inject multilingual semantic context early in the decoding process, thereby enhancing accuracy through adaptive cross-lingual guidance.

\section{Experimental Setup}

\subsection{Backbone Model}

Our framework is built upon Whisper \cite{radford2023}, an open-source speech recognition model developed by OpenAI.
Whisper has been trained on over 680,000 hours of multilingual and multi-task supervised data, enabling robust transcription across diverse accents, noise conditions, and domains.
The model is available in various sizes, from ``tiny'' to ``large,'' offering flexibility to balance computational efficiency and transcription accuracy.
For our experiments, we utilized the Whisper\textsubscript{Small} model, which provides an advantageous trade-off between recognition performance and resource requirements.
This makes it well suited for multilingual speech processing tasks in both research and applied settings.

\subsection{Model Configuration}

Model training was conducted in two stages to ensure stable convergence and effective adaptation to the target language.
In the first stage, the Whisper\textsubscript{Small} model was fine-tuned on Taiwanese Hokkien speech with a batch size of 4 and a learning rate of $1.25 \times 10^{-5}$ for 80,000 steps, including a warm-up phase of 8,000 steps.
The second stage resumed from the best checkpoint obtained in the first stage and continued training for 180,000 steps with a larger batch size of 8 and a learning rate of $5.0 \times 10^{-5}$, using 30,000 warm-up steps.
Both stages utilized the AdamW optimizer \cite{loshchilov2019} with a weight decay of 0.01 and employed mixed precision (FP16) to enhance training efficiency.
The audio input duration was constrained to 10 seconds, corresponding to a maximum of 160,000 samples, while the acoustic representation employed 80 mel-frequency bins.
For multilingual supervision, the number of auxiliary languages $L$ was set to 5, comprising Mandarin, English, Hindi, Spanish, and French, selected as the five most widely spoken languages globally to maximize linguistic diversity.
Both the auxiliary language embeddings and the acoustic representations were encoded with an embedding dimension of 768.

\subsection{Evaluation Metric}

Character error rate (CER) serves as the primary metric for evaluating the transcription accuracy of the proposed TG-ASR framework.
CER quantifies the proportion of incorrectly recognized characters by comparing the predicted transcription against the reference transcription.
This metric accounts for character substitutions, deletions, and insertions, offering a fine-grained measure of ASR performance.
CER is particularly appropriate for low-resource languages, such as Taiwanese Hokkien, where word boundaries can be ambiguous and tokenization standards may differ.
A lower CER signifies greater transcription accuracy, establishing it as a reliable metric for comparing various models and assessing advancements in speech recognition performance.
Note that the CERs reported in our experiments are computed under a teacher-forcing decoding setting to specifically evaluate the optimal integration of translation guidance.

\begin{table}[t]
\small
\centering
\setlength{\tabcolsep}{9.5pt}
\begin{tabular}{l|p{2.5cm}|cc}
\toprule
\bf ID & \bf Aux. Lang. & \bf CER \% & \bf Rel. \% \\
\toprule
A0 & - & 13.40 & - \\
\midrule
A1 & Mandarin (GT) & 11.87 & 11.42 \\
\midrule
A2 & Hindi & 13.17 & 1.72 \\
A3 & English & 13.10 & 2.24 \\
A4 & French & 12.98 & 3.13 \\
A5 & Spanish & 12.84 & 4.18 \\
\midrule
A6 & Mandarin (GT) \newline + Spanish & 11.42 & 14.77 \\
\bottomrule
\end{tabular}
% \vspace{-5pt}
\caption{
CERs and relative reductions (Rel.) on YT-THDC using different auxiliary languages (aux. lang.), where Mandarin serves as the ground-truth (GT) subtitle reference provided in the corpus.
}
\label{tab:main}
% \vspace{-5pt}
\end{table}

% \begin{table*}[t]
% \footnotesize % 調整為 footnotesize 以適應更多欄位
% \centering
% \setlength{\tabcolsep}{2pt} % 進一步縮減間距
% \begin{tabular}{l|l|c|*{11}{>{\centering\arraybackslash}p{2.2em}}} % 欄位改為 11 個
% \toprule
% \multirow{2}{*}{\textbf{ID}} & \multirow{2}{*}{\textbf{Auxiliary Language}} & \multirow{2}{*}{\textbf{Teacher Forcing}} & \multicolumn{11}{c}{\textbf{Autoregressive (Beam Size $b$)}} \\
% \cmidrule(lr){4-14} % 跨欄範圍更新為 4 到 14
% & & & \textbf{1} & \textbf{5} & \textbf{10} & \textbf{15} & \textbf{20} & \textbf{25} & \textbf{30} & \textbf{35} & \textbf{40} & \textbf{45} & \textbf{50} \\
% \midrule
% A0 & - & 13.40 & 22.24 & 21.32 & 21.18 & 21.16 & 21.16 & 21.20 & 21.20 & 21.23 & 21.19 & 21.18 & 21.17 \\
% \midrule
% A1 & Mandarin (GT) & 11.87 & 26.22 & 23.17 & 22.74 & 22.74 & 22.31 & 22.32 & 22.20 & 22.14 & 22.10 & 22.09 & 22.06 \\
% \midrule
% A2 & Hindi & 13.17 & - & - & - & - & - & - & - & - & - & - & - \\
% A3 & English & 13.10 & - & - & - & - & - & - & - & - & - & - & - \\
% A4 & French & 12.98 & - & - & - & - & - & - & - & - & - & - & - \\
% A5 & Spanish & 12.84 & - & - & - & - & - & - & - & - & - & - & - \\
% \midrule
% A6 & Mandarin (GT) + Spanish & 11.42 & - & - & - & - & - & - & - & - & - & - & - \\
% \bottomrule
% \end{tabular}
% \caption{CER comparison between Teacher Forcing (Oracle) and Autoregressive decoding with various beam sizes $b$.}
% \label{tab:beam_analysis}
% \end{table*}

\section{Results and Discussion}

\subsection{Main Results on YT-THDC}

Table~\ref{tab:main} presents the key findings from our experiments, which unequivocally illustrate the efficacy of the proposed translation-guided framework in improving ASR performance for Taiwanese Hokkien.
The baseline model (A0), trained exclusively on Taiwanese Hokkien transcripts without auxiliary supervision, serves as a reference point for performance evaluation.
Upon the introduction of auxiliary textual signals, consistent improvements are observed across all configurations, thereby validating the advantages of utilizing multilingual textual guidance.

Among single-language supervisors, the ground-truth Mandarin reference (A1) achieves the most significant reduction in CER.
This finding aligns with our hypothesis that a semantically close and high-quality auxiliary language offers the most effective guidance signal for model optimization.
The machine-translated auxiliary languages (A2 to A5) further demonstrate the robustness and generalizability of our framework.
Despite the presence of translation noise, all four languages lead to notable CER reductions compared to the baseline, indicating that semantic cues from cross-lingual text provide valuable supervision.
Interestingly, Spanish (A5) results in the most substantial improvement among the translated texts, surpassing typologically more distant languages such as Hindi (A2).
These results suggest that even approximate semantic alignment, as captured through machine translation, can provide significant benefits when utilized as auxiliary supervision.

The most significant outcome arises from the multilingual combination (A6), which attains the lowest CER and the highest relative improvement throughout the entire study, even surpassing the strong Mandarin-only condition (A1).
This finding indicates that the model adeptly leverages complementary information across languages, likely reaping the benefits of diverse linguistic viewpoints that strengthen shared semantics and mitigate overfitting to any individual translation source.
Collectively, these results provide compelling evidence in support of our hypothesis that the integration of multilingual textual information constitutes an effective and scalable approach for enhancing ASR systems in low-resource settings.

\begin{table}[t]
\small
\centering
\setlength{\tabcolsep}{14.5pt}
\begin{tabular}{l|l|c}
\toprule
\bf ID & \bf Configuration & \bf CER \% \\
\toprule
A6 & Full PGCA & 11.42 \\
\midrule
A7 & w/o tanh Gating & 11.46 \\
A8 & Sequential Attention & 11.60 \\
A9 & Shared Attention & 12.00 \\
\midrule
A10 & Addition & 27.68 \\
A11 & Concatenation & 24.09 \\
\bottomrule
\end{tabular}
% \vspace{-5pt}
\caption{Ablation study on the PGCA mechanism.}
\label{tab:ablation}
% \vspace{-5pt}
\end{table}

\subsection{Analytical Evaluation of PGCA}

To validate the effectiveness and architectural design of the proposed PGCA module, we performed a comprehensive ablation study.
The results, summarized in Table~\ref{tab:ablation}, systematically deconstruct the PGCA module to evaluate the contribution of each fundamental component.
Our complete model (A6), which incorporates all components, serves as the baseline for comparison.
We first analyzed the structural elements within the PGCA.
When the tanh gating mechanism was omitted (A7), performance exhibited a slight degradation, highlighting that the gating function is pivotal in regulating the flow of multilingual information.
Furthermore, substituting the original five-way parallel attention with a sequential configuration (A8) resulted in a higher CER, reinforcing the significance of enabling the decoder to attend to all auxiliary languages concurrently rather than sequentially.
Additionally, transforming the five independent attention branches into a shared-weight configuration (A9) further diminished accuracy, indicating that preserving independent attention parameters for each language facilitates more precise and language-specific interactions.

We further compared our proposed PGCA framework against two widely utilized fusion strategies.
Substituting the entire PGCA module with straightforward element-wise addition (A10) or concatenation (A11) led to significant performance degradation.
These results demonstrate that our PGCA module offers a more efficient approach for integrating multilingual representations through its synergistic use of parallel attention, independent weighting, and gated control.

\begin{table}[t]
\small
\centering
\setlength{\tabcolsep}{7.5pt}
\begin{tabular}{l|l|ccc}
\toprule
\multirow{2}{*}{\bf ID} & \multirow{2}{*}{\bf Aux. Lang.} & \multirow{2}{*}{\bf CER \%} & \multicolumn{2}{c}{\bf Proximity} \\
\cmidrule(lr){4-5}
 & & & \bf Tai. & \bf Man. \\
\toprule
A1 & Mandarin (GT) & 11.87 & 0.905 & - \\
A2 & Hindi & 13.17 & 0.854 & 0.879 \\
A3 & English & 13.10 & 0.552 & 0.590 \\
A4 & French & 12.98 & 0.821 & 0.847 \\
A5 & Spanish & 12.84 & 0.843 & 0.873 \\
\bottomrule
\end{tabular}
% \vspace{-5pt}
\caption{
CERs and language proximity between each auxiliary language translation and transcriptions of Taiwanese Hokkien (Tai.) subtitles of Mandarin (Man.) on YT-THDC.
}
\label{tab:similarity}
% \vspace{-5pt}
\end{table}

\subsection{Contribution of Auxiliary Languages}

To investigate why certain auxiliary languages yield greater improvements, we examined the relationship between ASR performance and language proximity.
We define language proximity in this context as the similarity between sentence representations derived from a pre-trained mBERT model, hypothesizing that this measure captures relevant linguistic and semantic factors beyond pure meaning alignment offered by models like LaBSE \cite{feng2022}.
Specifically, we measured this proximity using embeddings derived from the \texttt{[CLS]} token of the pre-trained mBERT model,hypothesizing that this might capture relevant linguistic and semantic factors.
For each sentence pair, we encoded the ground-truth Taiwanese Hokkien transcript and its machine-translated counterpart using mBERT, extracted their respective \texttt{[CLS]} embeddings, and computed the cosine similarity.
These scores were averaged across the dataset to yield a single proximity value between each auxiliary language and Taiwanese Hokkien (Tai.).
We also computed proximity relative to the Mandarin subtitles (Man.) present in our corpus.

Table~\ref{tab:similarity} presents the CERs alongside various proximity measures.
A discernible trend emerges when examining proximity to Taiwanese Hokkien: languages with higher proximity generally correlate with enhanced ASR performance; however, the relationship is not strictly monotonic.
Mandarin serves as the most prominent example, demonstrating the highest proximity and achieving the optimal CER. In contrast, English exhibits the lowest proximity and is associated with inferior ASR results.
Conversely, other languages exhibit more intricate patterns; for instance, Hindi, despite its high proximity, results in the worst CER, whereas Spanish outperforms French, even though both languages display similar proximity scores.

Notably, proximity scores computed in relation to the Mandarin subtitles exhibit a weaker correlation with the final performance of the Taiwanese Hokkien ASR.
This observation indicates that the direct proximity to the target language (Taiwanese Hokkien), as quantified by mBERT \texttt{[CLS]} similarity, serves as a more pertinent, albeit imperfect, predictor of an auxiliary language's potential contribution than does its proximity to the intermediate Mandarin text.
Moreover, other factors beyond this specific proximity measure are likely to impact the overall performance.

\begin{figure}[t]
\centering
\includegraphics[width=1.0\linewidth]{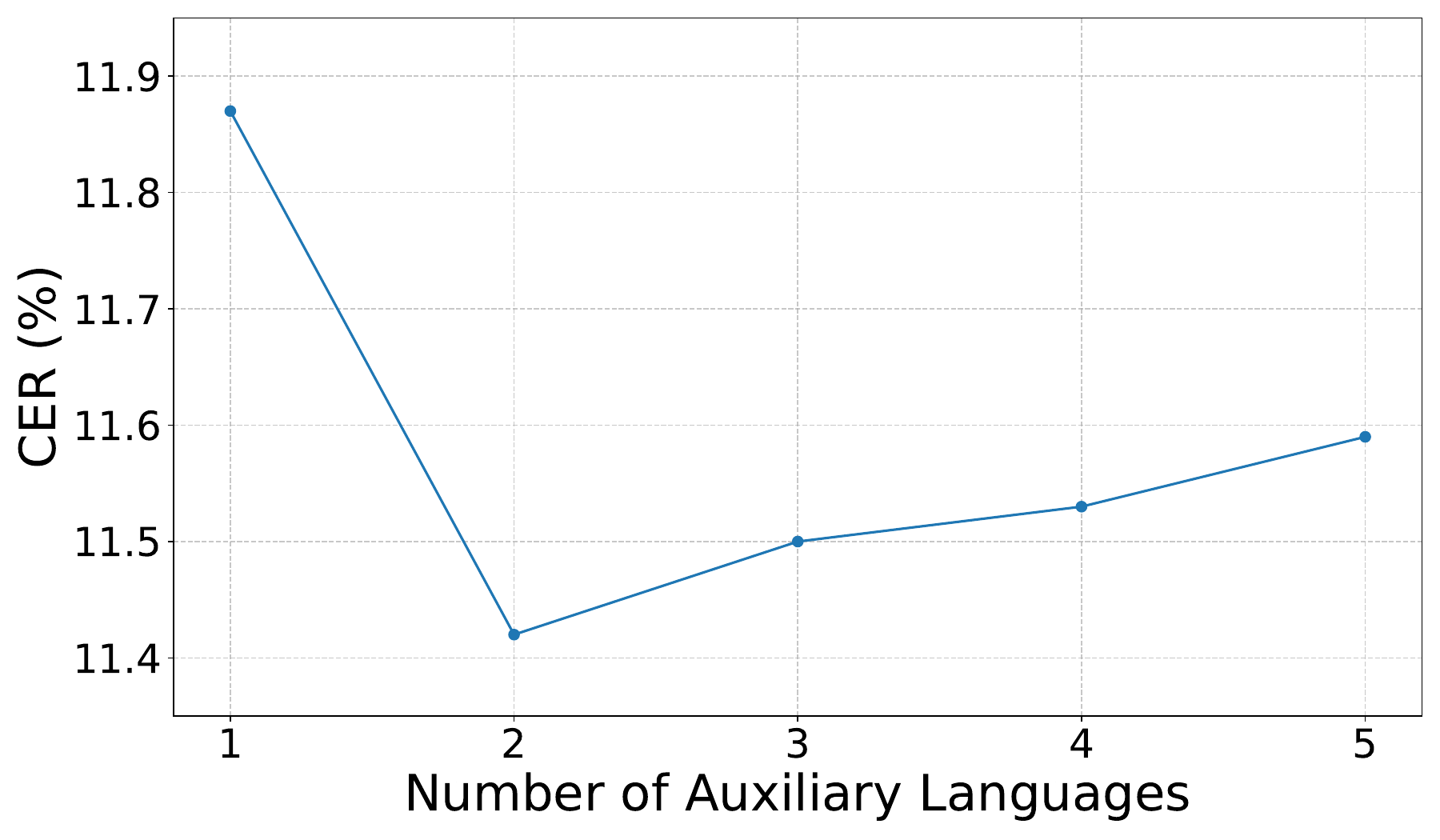}
% \vspace{-15pt}
\caption{CERs on YT-THDC as the number of auxiliary languages increases.}
% \vspace{-5pt}
\label{fig:scalability}
\end{figure}

\subsection{Analysis of Language Quantity}

To evaluate scalability and the impact of multilingual data quantity, we conducted an incremental experiment, cumulatively adding auxiliary languages based on their individual effectiveness (best first, see Table~\ref{tab:main}), up to all five languages.

Figure~\ref{fig:scalability} illustrates the outcomes, revealing a non-monotonic trend.
Performance significantly improves when adding the second-best language to the single best one, achieving the lowest CER.
This suggests a strong complementary effect between the top two auxiliary languages (Mandarin and Spanish).
However, progressively including more languages leads to a gradual CER increase, although performance remains substantially better than using only the single best language.
This indicates that adding languages with lower effectiveness or proximity may introduce some noise or interference that slightly diminishes the peak performance achieved with two languages.

Despite this slight upward trend after the second language, the results demonstrate the robustness of the PGCA mechanism.
It successfully leverages initial multilingual data for substantial gains and maintains strong performance even with less optimal languages, preventing significant degradation.
The learnable gating function likely plays a key role by adaptively weighting contributions, even if interference from the later-added languages is not perfectly suppressed.
This experiment highlights that while our framework benefits from multilingualism, carefully selecting the optimal number of auxiliary languages might yield the best results.

\begin{figure}[t]
\centering
\includegraphics[width=1.0\linewidth]{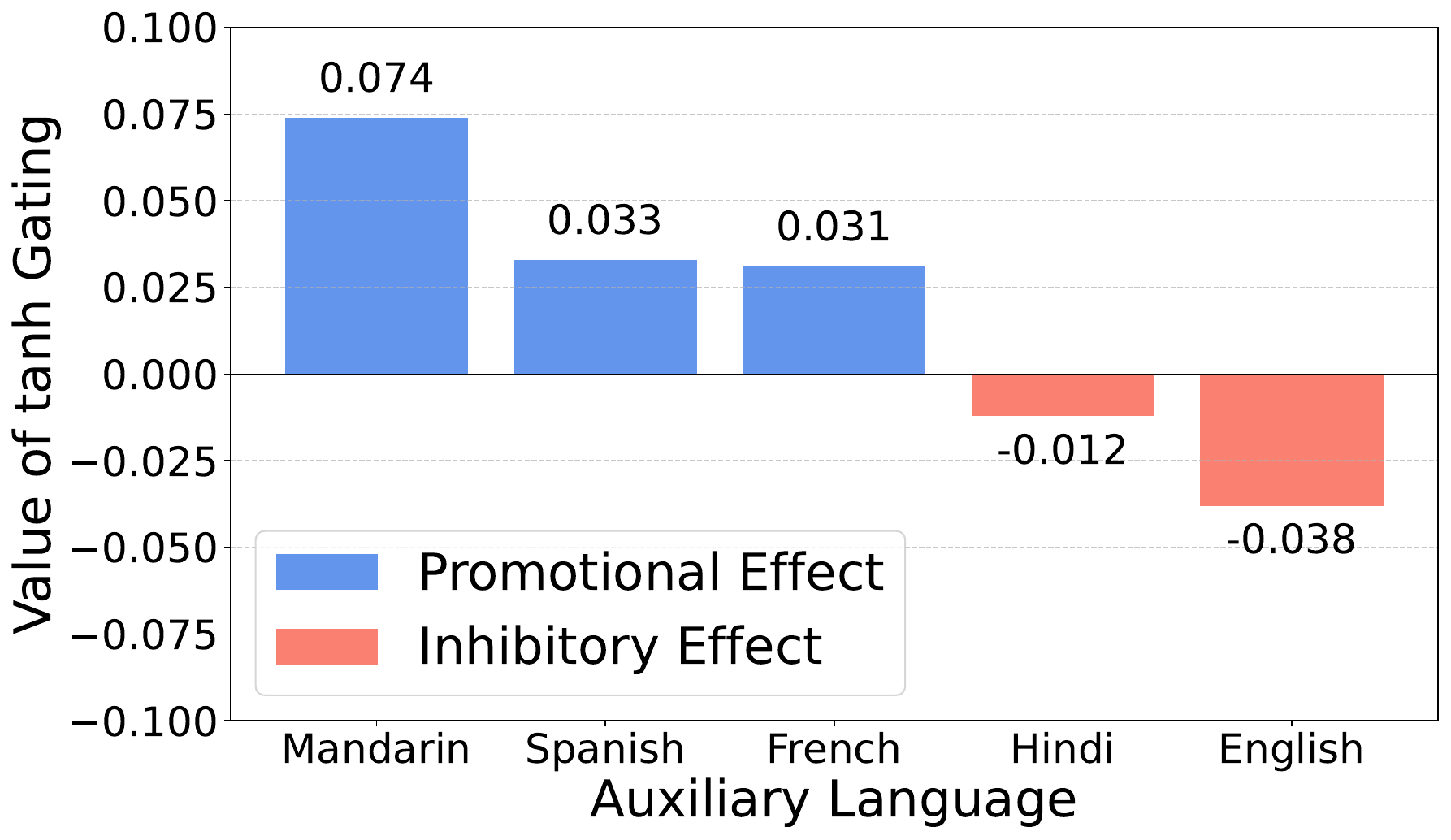}
% \vspace{-15pt}
\caption{Average activation values of the tanh gating mechanism across auxiliary languages, extracted from a representative decoder layer.}
% \vspace{-5pt}
\label{fig:weights}
\end{figure}

\subsection{Dynamic Behavior of tanh Gating}

To deepen our understanding of the proposed PGCA mechanism's regulation of multilingual information flow, we examined the functionality of its learnable tanh gating component. 
The underlying hypothesis posits that these gates are essential for dynamically balancing the contributions of various auxiliary languages, thus improving informative signals while attenuating potentially noisy or less relevant inputs.
To validate this hypothesis, we extracted the activation values of the tanh gates from a representative decoder layer of the trained model, where intricate semantic interactions are prominently established.

Figure~\ref{fig:weights} presents the averaged gating weights for each auxiliary language.
Positive activations correspond to languages that the model promotes during decoding, indicating a strong semantic alignment with the target speech.
Conversely, negative activations signify languages that the model suppresses, suggesting limited utility or potential interference in cross-lingual feature integration.
Mandarin, showcasing the highest semantic alignment with Taiwanese Hokkien, exhibits the strongest positive activation, corroborating that the gating mechanism adaptively prioritizes it as the most informative supervisory signal.
This behavior aligns with our main quantitative findings (see Table~\ref{tab:main}), where Mandarin supervision achieves the lowest CER.
In contrast, English and Hindi, which yield smaller performance gains in previous experiments, demonstrate negative gating activations, indicating that the model actively filters out less beneficial signals rather than passively ignoring them.

These observations underscore the adaptive and discriminative characteristics of the PGCA gating mechanism.
By learning to modulate the contributions of each auxiliary language, the model effectively achieves stability and robustness as multilingual inputs increase.
This dynamic control mechanism offers an internal rationale for the consistent performance improvements noted in our scalability analysis (see Figure~\ref{fig:scalability}), demonstrating that the model not only capitalizes on multilingual diversity but also develops an intelligent approach to managing it.

\begin{table}[t]
\small
\centering
\setlength{\tabcolsep}{18.5pt}
\begin{tabular}{l|l|c}
\toprule
\bf ID & \bf Model & \bf CER \% \\
\toprule
A6 & SeamlessM4T & 11.42 \\
A12 & NLLB & 11.52 \\
\bottomrule
\end{tabular}
% \vspace{-5pt}
\caption{Comparison of ASR performance using different translation models.}
\label{tab:comparison}
% \vspace{-5pt}
\end{table}

\subsection{Impact of Auxiliary Source}

To determine the optimal source of multilingual guidance for our TG-ASR framework, we conducted a comparative study between two leading multilingual translation models: SeamlessM4T \cite{barrault2023} and NLLB \cite{marta2022}.
The goal was to assess which model's generated translations provide more effective auxiliary supervision for Taiwanese Hokkien ASR.
For a fair evaluation, we employed the optimal two-language combination identified in our scalability analysis (Mandarin + Spanish) as the auxiliary input generated by each respective translation model.
% The impact of these translations on downstream ASR performance was systematically evaluated using our YT-THDC dataset.

The results are presented in Table~\ref{tab:comparison}.
The comparison indicates that both models function as potent sources for translation-guided learning; however, the incorporation of auxiliary texts generated by SeamlessM4T results in a markedly superior CER on our principal ASR task.
This performance advantage in speech recognition aligns with the reported intrinsic translation quality of SeamlessM4T , which demonstrated competitive overall text-to-text translation performance (chrF++) compared to NLLB on the standard multilingual FLORES \cite{goyal2022} benchmark.

This finding indicates that the quality of the auxiliary translations significantly influences the overall performance of the ASR system; a more accurate or semantically aligned translation is likely to offer a more robust guidance signal for the PGCA mechanism.
These results validate our choice of employing SeamlessM4T as the primary translation source for the main experiments conducted in this study.

\begin{figure}[t]
\centering
\includegraphics[width=1.0\linewidth]{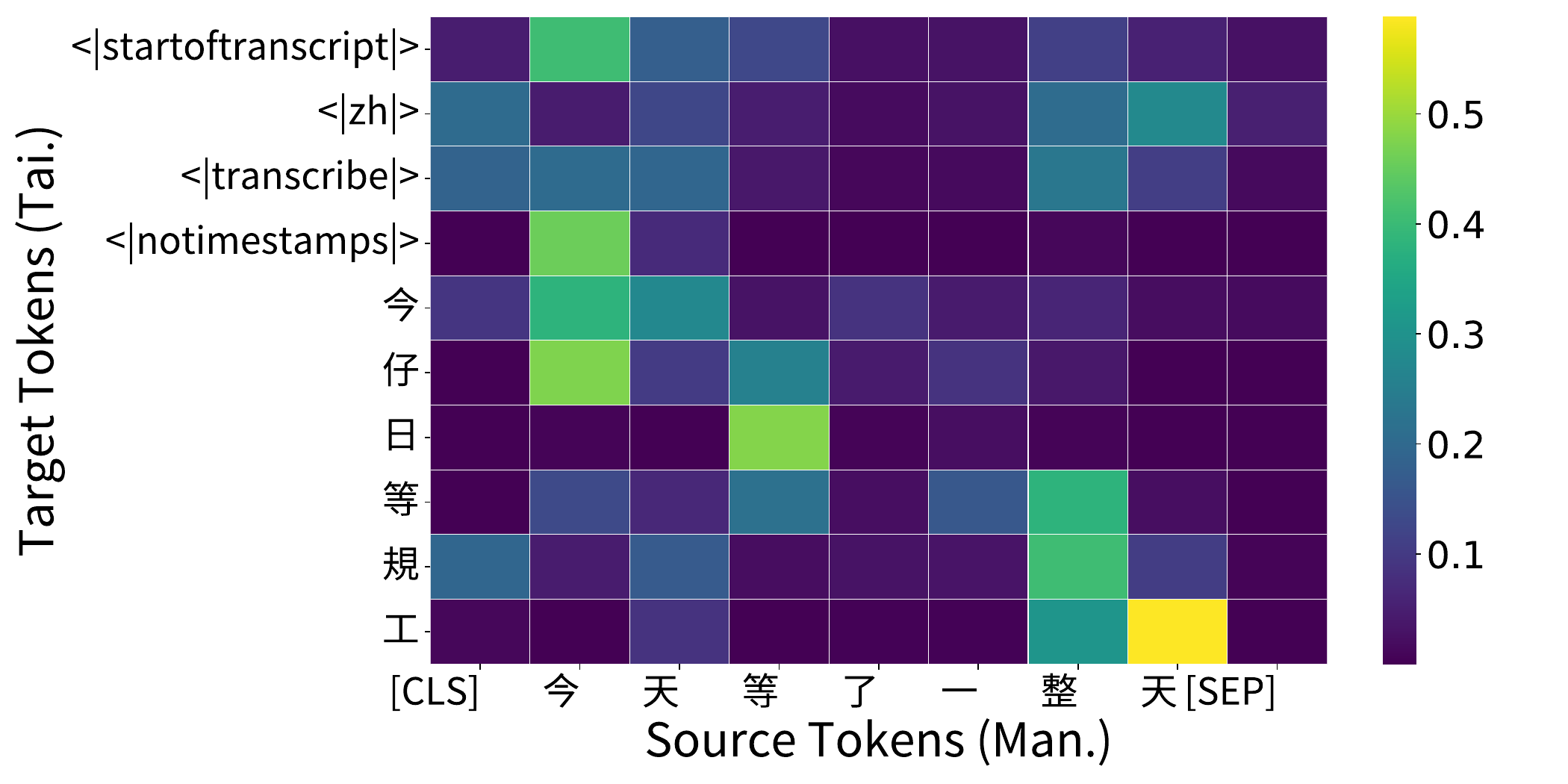}
% \vspace{-15pt}
\caption{
Visualization of the cross-lingual attention weights from a representative PGCA layer, mapping source Mandarin (Man.) tokens to target Taiwanese Hokkien (Tai.) tokens.
}
% \vspace{-5pt}
\label{fig:heatmap}
\end{figure}

\subsection{Cross-Lingual Attention}

To offer a qualitative and fine-grained analysis of our PGCA mechanism, we visualized the cross-attention weights to understand how the model aligns the source Mandarin translation with the generated Taiwanese Hokkien transcription at the token level.
We hypothesize that an effective translation-guided model should learn explicit correspondences between semantically related tokens across the two languages.
To test this, we extracted the attention matrix from a representative PGCA layer for a sample utterance from the test set.

\begin{CJK}{UTF8}{bkai}
Figure~\ref{fig:heatmap} presents the resulting heatmap, which provides compelling visual proof of our hypothesis.
A strong diagonal pattern emerges, indicating that the model successfully learns the token-level alignment between the two languages.
More impressively, the model demonstrates an understanding of semantic paraphrases beyond direct character matches.
It correctly aligns the Taiwanese Hokkien phrase ``今仔日'' (kin-á-ji̍t, today) with the Mandarin ``今天'' (jīntiān), and maps the Taiwanese Hokkien ``規工'' (kui-kang, whole day) to the Mandarin ``整天'' (zhěng tiān).
These highly focused attention weights confirm that the PGCA mechanism is not merely treating the auxiliary translation as a sentence-level feature vector or a ``bag of words.''
Instead, it is actively learning granular, cross-lingual semantic relationships.
This precise, token-level guidance provides a clean and unambiguous supervisory signal to the decoder, offering a clear explanation for the performance gains observed in our quantitative experiments.
\end{CJK}

\subsection{Analysis of Auxiliary Language Selection Strategies}

To determine the most effective method for selecting auxiliary languages under a constrained subset, we compared three data-driven strategies based on metrics derived from our prior analyses: individual language CER, language proximity, and learned gating values.
For each strategy, we identified the top-$k$ languages (where $k$ ranges from 1 to 5) based on ranking languages according to the respective metric (lower CER is better; higher proximity/gating value is better).
The resulting ASR performance for each strategy and value of $k$ is presented in Table~\ref{tab:strategy}.

At $k=1$, all three strategies identified Mandarin as the single best auxiliary language.
Consequently, they share the same CER result, establishing a common starting point.
However, divergences appear as more languages are added.
At $k=2$, selecting languages based on their individual CERs or their learned gating values yields the best performance, outperforming the strategy based purely on language Proximity.
This suggests that while proximity is important, the model's learned gating weights or the actual downstream task performance might be slightly better indicators for selecting the most complementary pair of languages.

Interestingly, at $k=3$, the strategies diverge, with the Proximity-based selection yielding slightly better performance than the CER-based and Gating Value-based selections.
At $k=4$, the trend shifts again, with selections based on proximity or gating value clearly outperforming the CER-based selection.
Finally, at $k=5$, all strategies naturally include all available languages, resulting in an identical language set and the same final CER.

Overall, these results indicate that while all three metrics provide reasonable heuristics for language selection, no single strategy consistently outperforms the others across all values of $k$.
Prioritizing based on CER or gating value is effective for smaller subsets ($k=2$), while proximity shows advantages at intermediate subset sizes ($k=3, 4$).
The analysis highlights that the optimal strategy might shift depending on the number of languages being combined.
Nevertheless, the differences between strategies remain relatively small, underscoring the effectiveness of the PGCA mechanism in managing diverse multilingual inputs selected through various data-driven approaches.

\begin{table}[t]
\small
\centering
\setlength{\tabcolsep}{4.5pt}
\begin{tabular}{l|ccccc}
\toprule
\multirow{2}{*}{\textbf{Strategy}} & \multicolumn{5}{c}{\textbf{\# Auxiliary Language ($k$)}} \\
\cmidrule(lr){2-6}
 & \textbf{1} & \textbf{2} & \textbf{3} & \textbf{4} & \textbf{5} \\
\midrule
CER & 11.87 & 11.42 & 11.50 & 11.53 & 11.59 \\ 
Proximity & 11.87 & 11.56 & 11.49 & 11.44 & 11.59 \\
Gating Value & 11.87 & 11.42 & 11.50 & 11.44 & 11.59 \\
\bottomrule
\end{tabular}
% \vspace{-5pt}
\caption{Comparison of ASR performance (CER \%) using different strategies for selecting the top-$k$ auxiliary languages.}
\label{tab:strategy}
% \vspace{-5pt}
\end{table}

% \begin{table}[t]
% \small
% \centering
% \setlength{\tabcolsep}{7.5pt}
% \begin{tabular}{l|l|ccc}
% \toprule
% \multirow{2}{*}{\bf ID} & \multirow{2}{*}{\bf Aux. Lang.} & \multirow{2}{*}{\bf CER \%} & \multicolumn{2}{c}{\bf Proximity} \\
% \cmidrule(lr){4-5}
%  & & & \bf Tai. & \bf Man. \\
% \toprule
% A1 & Mandarin (GT) & 11.87 & 0.905 & - \\
% A2 & Hindi & 13.17 & 0.854 & 0.879 \\
% A3 & English & 13.10 & 0.552 & 0.590 \\
% A4 & French & 12.98 & 0.821 & 0.847 \\
% A5 & Spanish & 12.84 & 0.843 & 0.873 \\
% \bottomrule
% \end{tabular}
% \vspace{-5pt}
% \caption{
% CERs and language proximity between each auxiliary language translation and transcriptions of Taiwanese Hokkien (Tai.)/subtitles of Mandarin (Man.) on YT-THDC.
% }
% \label{tab:similarity}
% \vspace{-10pt}
% \end{table}

\section{Conclusion and Future Work}

This study presents TG-ASR, a novel translation-guided framework tailored for Taiwanese Hokkien ASR, which effectively harnesses multilingual textual information to improve performance in scenarios with limited transcribed data.
Our key contributions encompass the development of the framework, which incorporates a novel parallel gated cross-attention mechanism, the release of the 30-hour YT-THDC corpus, and comprehensive experiments that reveal a substantial 14.77\% relative reduction in CER.
Analyses confirm that the PGCA module adaptively integrates diverse signals, effectively promoting beneficial languages while suppressing less informative ones.
These findings substantiate translation-guided learning as a powerful approach for improving ASR in practical, resource-constrained contexts.

For future work, we aim to address existing limitations and explore promising research directions.
A significant limitation is our reliance on auxiliary texts during inference; we are actively investigating knowledge distillation methodologies to develop a model that operates exclusively on speech input.
Additionally, we will examine the effects of translation quality and explore methodologies for enhancing robustness against noise present in machine-generated texts.
Lastly, applying TG-ASR to other underrepresented languages is essential for evaluating cross-lingual transferability and validating its effectiveness as a generalizable solution.

\section{Limitations}

First, YT-THDC is limited in size and domain, consisting of approximately 30 hours of Taiwanese Hokkien drama speech.
While it provides diverse speakers and recording conditions, the corpus does not cover other genres such as conversational speech, radio broadcasts, or spontaneous dialogues, which may limit the generalizability of TG-ASR to broader real-world scenarios.
Second, the TG-ASR framework relies on auxiliary translations generated by pre-trained multilingual models.
Translation errors, misalignments, or semantic distortions may introduce noise into the cross-lingual supervision, potentially affecting model performance.
Although the PGCA mechanism mitigates some interference, extremely low-quality translations or languages with low language proximity may provide limited benefits.
Third, TG-ASR is evaluated exclusively on Taiwanese Hokkien as the target language, and the observed improvements may vary for other low-resource languages with different phonological, syntactic, or morphological characteristics.
Future work is needed to assess cross-lingual transferability, domain adaptation, and scalability to larger and more diverse datasets.

\section{Ethical Considerations}

The YT-THDC corpus is constructed from publicly available Taiwanese Hokkien drama videos on YouTube, sourced from official broadcasting channels.
Only non-commercial research purposes are intended, and no personally identifiable information beyond what is publicly visible has been collected.
All manual transcriptions were performed by trained linguistic experts under privacy-respecting procedures.
The use of machine-translated auxiliary language data may introduce biases, including translation errors or semantic misalignments, which could affect model behavior.
Outputs from the TG-ASR framework should be treated as assistive rather than authoritative, especially in contexts requiring high transcription accuracy.
YT-THDC is released under terms restricting its use to non-commercial research and educational purposes. 
Potential societal impacts, including benefits for low-resource language research and risks of propagating errors, are carefully considered to encourage responsible use and development.

\section{Bibliographical References}

\bibliographystyle{lrec2026-natbib}
\bibliography{references.bib}
\end{document}